\documentclass[twoside,11pt]{article}

\usepackage{jmlr2e}
\usepackage{float}
\usepackage{tabularx}
\usepackage{caption}
\usepackage{algorithm}
\usepackage{algorithmic}
\usepackage[utf8]{inputenc}
\usepackage{amsmath}
\usepackage{amsfonts}

\jmlrheading{23}{2023}{1-10}{6/25}{00/00}{Suman Kunwar}

\ShortHeadings{Convolutional Neural Network (CNN) to reduce construction loss in JPEG compression}{Kunwar}
\firstpageno{1}

\begin{document}

\title{Convolutional Neural Network (CNN) to reduce construction loss in JPEG compression caused by Discrete Fourier Transform (DFT)}

\author{\name Suman Kunwar \email sumn2u@gmail.com \\
       \addr Faculty of Computer Science\\
       Selinus University of Sciences and Literature\\
       Ragusa, Italy
       }


\maketitle

\begin{abstract}
In recent decades, digital image processing has gained enormous popularity.  Consequently, a number of data compression strategies have been put forth, with the goal of minimizing the amount of information required to represent images. Among them, JPEG compression is one of the most popular methods that has been widely applied in multimedia and digital applications. The periodic nature of DFT makes it impossible to meet the periodic condition of an image's opposing edges without producing severe artifacts, which lowers the image's perceptual visual quality. On the other hand, deep learning has recently achieved outstanding results for applications like speech recognition, image reduction, and natural language processing. Convolutional Neural Networks (CNN) have received more attention than most other types of deep neural networks. The use of convolution in feature extraction results in a less redundant feature map and a smaller dataset, both of which are crucial for image compression. In this work, an effective image compression method is purposed using autoencoders. The study's findings revealed a number of important trends that suggested better reconstruction along with good compression can be achieved using autoencoders.
\end{abstract}


\section{Introduction}

In today’s digital era, human beings are surrounded by digital gadgets. Photographs are now an integral part of a person's daily life, and digital images are widely used in a variety of applications. As digital imaging and multimedia services advance, more and more people can share their data on the Internet. The number of internet users is growing day by day rapidly~\citep{rahman_histogram_2018}, resulting in increased data transfer, which necessitates efficient image compression. In multimedia, JPEG is one of the most commonly used lossy compression techniques~\citep{Hussain_2020}. There are several variations of JPEG: JPEG 2000, JPEG XS, JPEG Systems, JPEG Pleno, and JPEG XL~\citep{jpeg_future_2020}. Based on the data provided by Web Technology Survey, 74.3\% of websites use the JPEG image format~\citep{w3techs_usage_2022}. In digital images, pixels  have high correlations, and the removal of this correlation will not affect the visual quality of the image~\citep{gonzalez2009digital, yuan2019research}. To achieve the best quality with the smallest possible size, the low frequency values are preserved as they define the content of the image, and the high frequency values are truncated by a certain amount~\citep{li2019joint,rasheed_image_2020}.

With the help of DFT, images in the spatial domain can be converted into the frequency domain, and certain frequencies can be ignored or modified to produce a low-information image with adequate quality~\citep{siddeq2013applied, siddeq2014new,siddeq2014novel}. In practical application, when we compute the DFT of an image, it is impossible to meet the periodic condition that opposite borders of an image are alike, and the image always shows strong discontinuities across the frame border. As a result this affects the registration accuracy and success rate~\citep{dong_eliminating_2019}. To solve this problem, various approaches have been taken. Among them, raised-cosine window~\citep{leprince2007automatic}, blackman window~\citep{podder_comparative_2014}, and flap-top window~\citep{GE20146709} are the most popular ones.\newline
\hspace*{1.5em}As CNN have advanced rapidly~\citep{pouyanfar_survey_2019}, the problem of removing image artifacts from the decoded images has been re-examined. Several state-of-the-art deep learning-based algorithms~\citep{li_2020,svoboda_compression_2016,baig_learning_2017,santurkar_generative_2017,dong_eliminating_2019} have been developed with great success using this approach. Approaches like using the Hann windows for reducing edge-effects in patch based image segmentation with CNNs has shown promising results and pointed out the further investigation with different window functions, and with reducing the amount of context needed~\citep{pielawski_introducing_2020}. While these solutions are promising, still some information is lost in the process.\newline
\hspace*{1.5em} 
To minimize the reconstruction loss, an autoencoder network is purposed. The encoder reduces the
dimensionality of an image into a number of feature maps. The feature map contains densely packed color values of an image whose original value can be closely reconstructed using the decoder associated with it.  To measure the quality of the reconstructed image Mean Squared Error (MSE), Peak Signal to Noise Ratio (PSNR), and Structural Similarity Index (SSMI) are used. It is also compare with the popular image compression standard JPEG. Here, SSIM measures the similarity in pixels and is used as a second image quality metric. 
\newline
\hspace*{1.5em}
The remainder of this paper is organized as follows. Proposed method is described in Section II. In Sections III and IV, training and experimental evaluations are described in detail. Section V presents the conclusions of this research.

\section{Methods}
This section gives more information about the implementation, including deep learning network architecture, and the loss function.

\subsection{Network Architecture}
\begin{figure}
\centering
\includegraphics[width=0.9\textwidth, height=7.0cm]{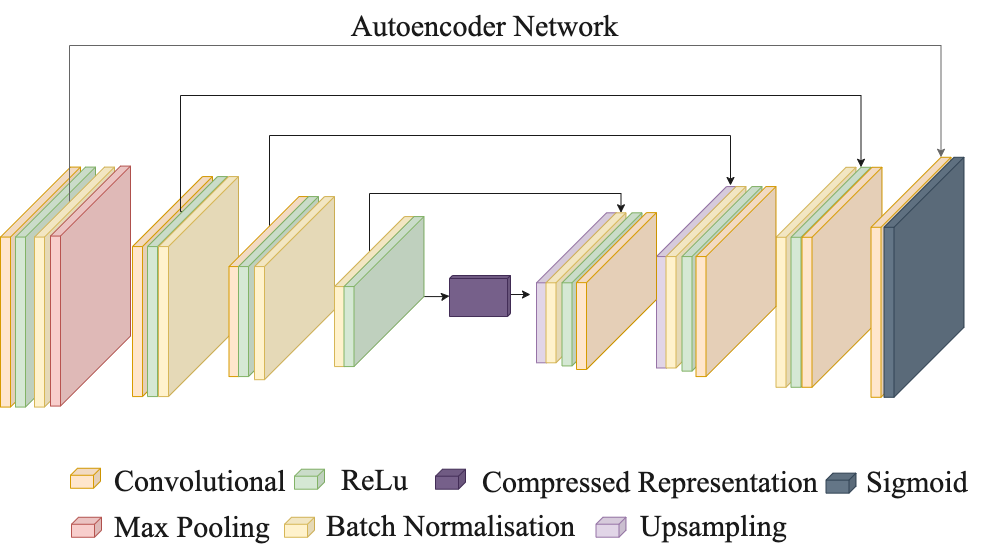}
 \caption{Proposed Network Architecture that consists of different level of convolutional layer}
 \label{fig:proposed_compression_tech}
\end{figure}

 As autoencoder pipeline claims to provide efficient compression efficiency~\citep{8456308, arxiv.1902.07385}, an autoenoder network is constructed. Here, the encoder takes the input to represent the data in the simplest possible way. This is done by extracting the most prominent features and presenting them in a way that the decoder can understand. Whereas, the decoder learns to read compressed code representations and generate them based on those representations. The proposed network architecture is depicted  in \autoref{fig:proposed_compression_tech}.

The encoders are trained along with the decoders without labels. Eight layer of convolution is used in the auto encoder network. Batch Normalization is used to enable
faster and more stable training of deep neural networks~\citep{arxiv.1805.11604}. The benefit of using a layer is, it allows similar operations to be performed simultaneously. With more convolution kernels, the number of parameters increases linearly. As a result, the number of output channels also increases linearly and helps to reduce the artifacts caused by the gaussian noise~\citep{audhkhasi_noise-enhanced_2016}. The computation time is also proportional to the size of the input channel and to the number of kernels.\\
  
Sigmoid  activation function  is used in the output layer, whose output is bound  between  0 and 1 range, and can be prone to suffering from the vanishing gradient problem~\citep{arxiv.2204.02921}. To overcome this, Rectified Linear Unit (ReLU) is  used as an activation function to increase to speed up the application and for better results~\citep{arxiv.2109.14545}. Upsampling is also done to increase the spatial dimensions of the feature maps~\citep{arxiv.2012.09904}. Adam optimizer is used as optimizer which uses momentum and adaptive gradient to compute adaptive learning rates for each parameter~\citep{arxiv.1412.6980}.
\begin{algorithm}[H]
\caption{Compression Model}
\begin{algorithmic}[1]

\STATE Build the compression model:
\STATE \quad \textbf{def} build\_compression\_model(img\_shape):
\STATE \quad \quad $encoder \gets \text{Sequential}()$
\STATE \quad \quad $encoder.\text{add}(\text{Conv2D}(32, kernel\_size=3, strides=1, padding='same', activation='relu', input\_shape=img\_shape))$ \COMMENT{32x32x32}
\STATE \quad \quad $encoder.\text{add}(\text{BatchNormalization}())$ \COMMENT{32x32x32}
\STATE \quad \quad $encoder.\text{add}(\text{MaxPooling2D}(2, padding='same'))$ \COMMENT{16x16x32}
\STATE \quad \quad $encoder.\text{add}(\text{Conv2D}(16, kernel\_size=3, strides=1, padding='same', activation='relu'))$ \COMMENT{16x16x16}
\STATE \quad \quad $encoder.\text{add}(\text{BatchNormalization}())$ \COMMENT{16x16x16}
\STATE \quad \quad $encoder.\text{add}(\text{Conv2D}(8, kernel\_size=3, strides=1, padding='same', activation='relu'))$ \COMMENT{16x16x8}
\STATE \quad \quad $encoder.\text{add}(\text{BatchNormalization}())$ \COMMENT{16x16x8}
\STATE \quad \quad $encoder.\text{add}(\text{Conv2D}(8, kernel\_size=3, strides=1, padding='same', activation='relu'))$ \COMMENT{16x16x8}

\STATE \quad \quad $decoder \gets \text{Sequential}()$
\STATE \quad \quad $decoder.\text{add}(\text{Conv2D}(32, kernel\_size=3, strides=1, padding='same', activation='relu'))$ \COMMENT{16x16x32}
\STATE \quad \quad $decoder.\text{add}(\text{BatchNormalization}())$ \COMMENT{16x16x32}
\STATE \quad \quad $decoder.\text{add}(\text{UpSampling2D}())$ \COMMENT{32x32x32}
\STATE \quad \quad $decoder.\text{add}(\text{Conv2D}(16, kernel\_size=3, strides=2, padding='same', activation='relu'))$ \COMMENT{16x16x16}
\STATE \quad \quad $decoder.\text{add}(\text{BatchNormalization}())$ \COMMENT{16x16x16}
\STATE \quad \quad $decoder.\text{add}(\text{UpSampling2D}())$ \COMMENT{32x32x16}
\STATE \quad \quad $decoder.\text{add}(\text{Conv2D}(16, kernel\_size=3, strides=1, padding='same', activation='relu'))$ \COMMENT{32x32x16}
\STATE \quad \quad $decoder.\text{add}(\text{BatchNormalization}())$ \COMMENT{32x32x16}
\STATE \quad \quad $decoder.\text{add}(\text{Conv2D}(3, kernel\_size=1, strides=1, padding='same', activation='sigmoid'))$ \COMMENT{32x32x3}

\STATE \quad \quad \textbf{return} encoder, decoder

\end{algorithmic}
\end{algorithm}

\begin{algorithm}[H]
\caption{Model Training}
\begin{algorithmic}[1]
\STATE Load CIFAR-10 dataset: $(X_{\text{train}}, Y_{\text{train}}), (X_{\text{test}}, Y_{\text{test}}) \gets \text{cifar10.load\_data()}$

\STATE Normalize the input data:
\STATE \quad $X_{\text{train}} \gets \frac{X_{\text{train}}}{255}$
\STATE \quad $X_{\text{test}} \gets \frac{X_{\text{test}}}{255}$
\STATE \quad $Y_{\text{test}} \gets \frac{Y_{\text{test}}}{255}$

\STATE Determine image dimensions:
\STATE \quad $img\_rows \gets X_{\text{train}}[0].\text{shape}[0]$
\STATE \quad $img\_cols \gets X_{\text{test}}[0].\text{shape}[1]$

\STATE Reshape training and test datasets:
\STATE \quad $X_{\text{train}} \gets X_{\text{train}}.\text{reshape}(\text{len}(X_{\text{train}}), X_{\text{train}}.\text{shape}[1], X_{\text{train}}.\text{shape}[2], 3)$

\STATE Define image input shape:
\STATE \quad $\text{IMG\_SHAPE} \gets X_{\text{train}}.\text{shape}[1:]$
\STATE \quad $input\_img \gets \text{Input(shape=IMG\_SHAPE)}$

\STATE \quad $encoder, decoder \gets \text{build\_compression\_model(IMG\_SHAPE)}$
\STATE \quad $code \gets \text{encoder}(input\_img)$
\STATE \quad $reconstruction \gets \text{decoder}(code)$
\STATE \quad $autoencoder \gets \text{Model}(input\_img, reconstruction)$

\STATE Compile the compression model:
\STATE \quad $\text{compression model}.\text{compile}(\text{optimizer}='adam', \text{metrics}=['accuracy'], \text{loss}='mean\_squared\_error')$

\STATE Set up early stopping and model checkpoint:
\STATE \quad $es \gets \text{EarlyStopping}(\text{monitor}='val\_loss', \text{mode}='min', \text{patience}=200)$
\STATE \quad $mc \gets \text{ModelCheckpoint}('auto\_encoder\_model.h5', \text{monitor}='val\_accuracy', \text{mode}='max', \text{verbose}=1, \text{save\_best\_only}=\text{True})$

\STATE Train the compression  model:
\STATE \quad $\text{compression model}.\text{fit}(x=X_{\text{train}}, y=X_{\text{train}}, \text{validation\_data}=[X_{\text{test}}, X_{\text{test}}], \text{epochs}=200, \text{callbacks}=[es, mc])$
\end{algorithmic}
\end{algorithm}

\subsection{Loss Function}
Loss functions are used to calculate predicted errors created across the training samples. Here, loss is measured in terms of Mean Squared Error (MSE). It calculates the mean of square error of the predicted value and the actual value. It results are derivable and is possible to control the update rate. 
\begin{equation}
\sum_{i=1}^{D}(x_i-y_i)^2
\end{equation}

\section{Training}
This section describes the training procedure and parameter settings used in this research.
\subsection{Datasets}
 CIFAR-10 dataset is used here to train and test the model~\citep{shorten_survey_2019}. It consists of 60000 32x32 colour images and is divided into training and test datasets. Simple random sampling is used here to represent the different sets of images as it gives an equal probability of selecting a particular item~\citep{9543483}. Here, samples are only taken from the training images, the dataset used to train the algorithm. As the method aims to evaluate itself using both real and representative data, no training were done on holdout tests~\citep{7544814}.

\subsection{Procedure}
The training dataset is run through the model in an incremental fashion. While training the model, Adam is used as a optimizer and  loss is calculated in terms of Mean Square Error (MSE) as it has better generalization performances than the Cross Entropy (CE) loss~\citep{arxiv.2006.07322}. Loss values and accuracy scores while training at different iterations are measured. For error correction, regularization options~\citep{arxiv.2110.11402} such as early stopping and dropout are carried out.

\section{Evaluation}
\begin{figure}[H]
\centering
\includegraphics[width=0.95\textwidth,height=6cm]{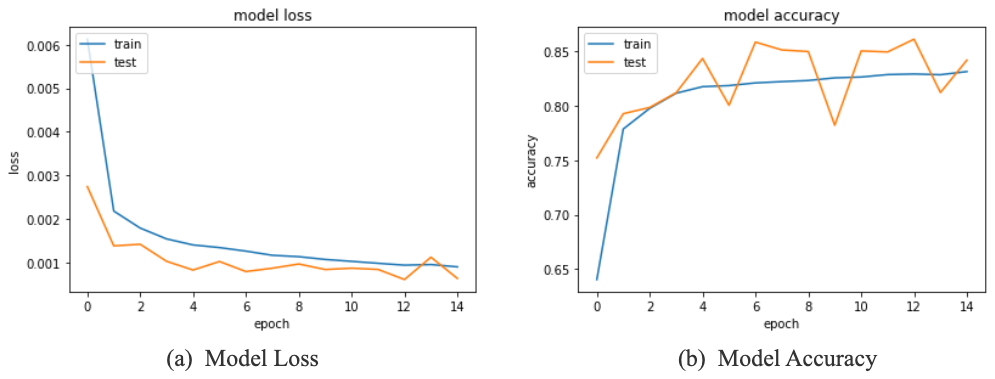}
\caption{Loss vs  Model Loss (a) and Accuracy vs Model Accuracy after 15 epoches of training}
 \label{fig:loss_vs_model_loss_15}
\end{figure}

This section compares the  results that happened while training the model with different settings. First, the model is tested with lower epochs as shown in \autoref{fig:loss_vs_model_loss_15} and then gradually increased.

\begin{figure}[H]
\centering
\includegraphics[width=0.90\textwidth,height=6cm]{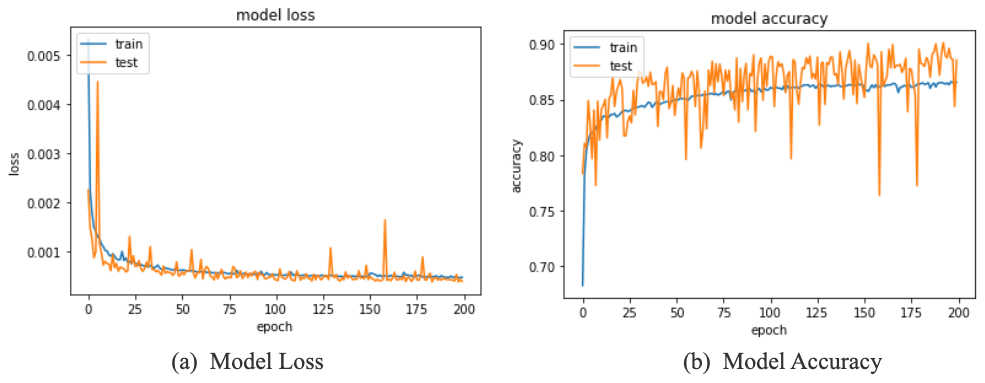}
\caption{Loss vs  Model Loss (a) and Loss vs  Model Accuracy (b) after 200 epoches of training}
 \label{fig:loss_vs_model_loss_200}
\end{figure}

\vspace{1em}
Certain spikes can be seen in \autoref{fig:loss_vs_model_loss_200} (a). Here, the loss first decreases, increases, and decreases at last. These spikes are often encountered when training with high learning rates, high order loss functions
or small batch sizes~\citep{Ede_2020}. The batch size is reduced and early stopping is added to the model for to avoid overfitting ~\citep{arxiv.2002.11569}. A checkpoint is also added to save the best value weights obtained while training. The maximum accuracy value has reached up to 0.90104 within 200 epochs training.

From the comparison shown in \autoref{tab:model_results_variations_comparison}, the model shows high accuracy with regularization and has higher loss with low training. With the increasing number of training variation on loss and accuracy can be seen. The model shows low accuracy with the addition of noise value to the model with normalization.

\begin{table}[H]
  \begin{center}
   \caption{Comparison of proposed model with different settings}
    \label{tab:model_results_variations_comparison}
    \renewcommand{\arraystretch}{1.5}
     \begin{tabularx}{\textwidth}{@{} X c c r @{}}
    \hline
    \textbf{Model (variations)} & \textbf{Loss} & \textbf{Accuracy} & \textbf{Epoches} \\
    \hline
       Model & 0.162 & 80.236 & 15 \\
       Model & 0.092 & 85.352 & 100 \\
       Model & 0.089 & 82.307 & 200 \\
       Model with batch size 32 & 0.13 & 84.269 & 200 \\
       Model with normalization & 0.07 & 85.94 & 100 \\
       Model with normalization & 0.08 & 87.22 & 200 \\
        Model with normalization and noise & 0.111 & 56.74 & 30 \\
    \hline
  \end{tabularx}
  \end{center}
 
\end{table}

Image quality can be compromised as a result of distortions during the acquisition and processing of images. Different metrics have been used to measure the quality of compressed results, including Mean Square Error (MSE), Peak Signal to Noise Ratio (PSNR), and Structural Similarity Index (SSMI)~\citep{deshmukh_image_2019_1}. To measure the performance of the reconstructed images of the purposed system, PSNR and SSIM are used.

\begin{table}[H]
  \begin{center}
    \caption{Performance comparison of original and DFT output}
    \label{tab:model_results_comparison_dft}
     \renewcommand{\arraystretch}{1.5}
     \begin{tabularx}{\textwidth}{@{} X c c c r @{}}
    \hline
      \textbf{Compression Ratio} &\textbf{MSE} &\textbf{PSNR} &\textbf{SSIM}  &\textbf{Size} \\
      \hline 90 & 0.000194 & 37.102 & 0.982 & 2.613 KB
      \\
       80 & 0.000267 & 35.722 & 0.97 & 2.602 KB 
      \\
      70 & 0.00037 & 34.309 & 0.96 & 2.599 KB 
      \\
      60 & 0.000521 & 32.827 & 0.956 & 2.577 KB 
      \\
       50 & 0.000731 & 31.356 & 0.938 & 2.560 KB \\
        \hline
    \end{tabularx}
    
  \end{center}
\end{table}

\begin{table}[H]
  \begin{center}

    \caption{Size comparison of original and model output image}
    \label{tab:size_comparison_original_model}
    \renewcommand{\arraystretch}{1.5}
  \begin{tabularx}{\textwidth}{@{} X c r @{}}
    \hline
    \textbf{} &\textbf{Original Image} &\textbf{Model Output} \\
    \hline
     Size & 2.594KB  & 2.543 KB \\
    \hline
  \end{tabularx}
   
    \end{center}
\end{table}

\begin{table}[H]
  \begin{center}
  \caption{Performance comparison of original and model output}
    \label{tab:model_results_comparison_original}
    \renewcommand{\arraystretch}{1.5}
     \begin{tabularx}{\textwidth}{@{} X c c r @{}}
    \hline
    \textbf{} & \textbf{MSE} &\textbf{PSNR} &\textbf{SSIM} \\
    \hline
      Values & 0.000199 & 37.0081  & 0.986\\
    \hline
  \end{tabularx}
  \end{center}
\end{table}

\begin{figure}[H]
\centering
\includegraphics[width=0.9\textwidth]{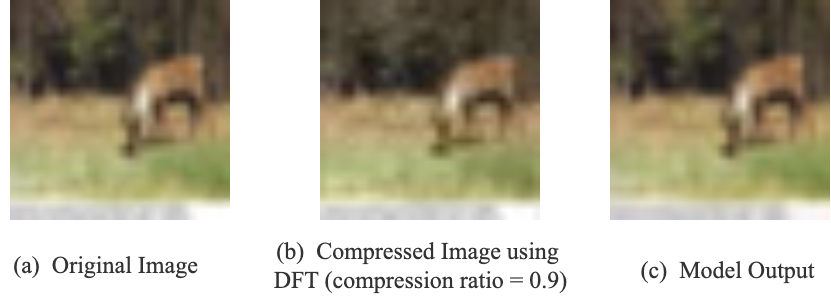}
\caption{Original image (a) and the reconstructed image produced by (b) DFT and the (c) purposed model}
 \label{fig:image_outputs}
\end{figure}

The results from  \autoref{tab:model_results_comparison_dft}, \autoref{tab:size_comparison_original_model}, and \autoref{tab:model_results_comparison_original} shows that the purposed system has better compression while preserving the quality. \autoref{fig:image_outputs} presents few sample images to compare the subjective quality of these methods. 
\vspace{1em}

\section{Conclusion}

An autoencoder-based image compression technique is presented in the paper, which uses a lower-dimensional representation to reconstruct the image with less reconstruction loss by capturing the most significant elements. While it outperforms  JPEG  compression using DFT, there are still other variations of JPEG, that needs further research. Among others, the impact of noise as a regularization method deserves further investigation. In our case, the addition of noise resulted in moderate quality images.

\vfill

{\noindent \em See \href{https://github.com/sumn2u/neuralnetwork-jpeg}{https://github.com/sumn2u/neuralnetwork-jpeg} for source code.}
\newpage

\vskip 0.2in
\bibliography{sample}

\begin{thebibliography}{37}
\providecommand{\natexlab}[1]{#1}
\providecommand{\url}[1]{\texttt{#1}}
\expandafter\ifx\csname urlstyle\endcsname\relax
  \providecommand{\doi}[1]{doi: #1}\else
  \providecommand{\doi}{doi: \begingroup \urlstyle{rm}\Url}\fi

\bibitem[Alexandre et~al.(2019)Alexandre, Chang, Peng, and
  Hang]{arxiv.1902.07385}
David Alexandre, Chih-Peng Chang, Wen-Hsiao Peng, and Hsueh-Ming Hang.
\newblock An autoencoder-based learned image compressor: Description of
  challenge proposal by nctu, 2019.
\newblock URL \url{https://arxiv.org/abs/1902.07385}.

\bibitem[Audhkhasi et~al.(2016)Audhkhasi, Osoba, and
  Kosko]{audhkhasi_noise-enhanced_2016}
Kartik Audhkhasi, Osonde Osoba, and Bart Kosko.
\newblock Noise-enhanced convolutional neural networks.
\newblock \emph{Neural Networks}, 78:\penalty0 15--23, June 2016.
\newblock ISSN 08936080.
\newblock \doi{10.1016/j.neunet.2015.09.014}.
\newblock URL
  \url{https://linkinghub.elsevier.com/retrieve/pii/S0893608015001896}.

\bibitem[Baig et~al.(2017)Baig, Koltun, and Torresani]{baig_learning_2017}
Mohammad~Haris Baig, Vladlen Koltun, and Lorenzo Torresani.
\newblock Learning to inpaint for image compression.
\newblock \emph{arXiv:1709.08855 [cs]}, November 2017.
\newblock URL \url{http://arxiv.org/abs/1709.08855}.
\newblock arXiv: 1709.08855.

\bibitem[Cheng et~al.(2018)Cheng, Sun, Takeuchi, and Katto]{8456308}
Zhengxue Cheng, Heming Sun, Masaru Takeuchi, and Jiro Katto.
\newblock Deep convolutional autoencoder-based lossy image compression.
\newblock In \emph{2018 Picture Coding Symposium (PCS)}, pages 253--257, 2018.
\newblock \doi{10.1109/PCS.2018.8456308}.

\bibitem[Deshmukh(2019)]{deshmukh_image_2019_1}
Kunal~Rajan Deshmukh.
\newblock \emph{Image compression using neural networks}.
\newblock Master of {Science}, San Jose State University, San Jose, CA, USA,
  May 2019.
\newblock URL \url{https://scholarworks.sjsu.edu/etd_projects/666}.

\bibitem[Dong et~al.(2019)Dong, Jiao, Long, Liu, and He]{dong_eliminating_2019}
Yunyun Dong, Weili Jiao, Tengfei Long, Lanfa Liu, and Guojin He.
\newblock Eliminating the effect of image border with image periodic
  decomposition for phase correlation based remote sensing image registration.
\newblock \emph{Sensors}, 19\penalty0 (10):\penalty0 2329, May 2019.
\newblock ISSN 1424-8220.
\newblock \doi{10.3390/s19102329}.
\newblock URL \url{https://www.mdpi.com/1424-8220/19/10/2329}.

\bibitem[Dubey et~al.(2021)Dubey, Singh, and Chaudhuri]{arxiv.2109.14545}
Shiv~Ram Dubey, Satish~Kumar Singh, and Bidyut~Baran Chaudhuri.
\newblock Activation functions in deep learning: A comprehensive survey and
  benchmark, 2021.
\newblock URL \url{https://arxiv.org/abs/2109.14545}.

\bibitem[Ede and Beanland(2020)]{Ede_2020}
Jeffrey~M Ede and Richard Beanland.
\newblock Adaptive learning rate clipping stabilizes learning.
\newblock \emph{Machine Learning: Science and Technology}, 1\penalty0
  (1):\penalty0 015011, mar 2020.
\newblock \doi{10.1088/2632-2153/ab81e2}.
\newblock URL \url{https://doi.org/10.1088%2F2632-2153%2Fab81e2}.

\bibitem[Ge et~al.(2014)Ge, Lan, and Wang]{GE20146709}
Peng Ge, Caiyu Lan, and Hong Wang.
\newblock An improvement of image registration based on phase correlation.
\newblock \emph{Optik}, 125\penalty0 (22):\penalty0 6709--6712, 2014.
\newblock ISSN 0030-4026.
\newblock \doi{https://doi.org/10.1016/j.ijleo.2014.07.086}.
\newblock URL
  \url{https://www.sciencedirect.com/science/article/pii/S0030402614010377}.

\bibitem[Gonzalez et~al.(2009)Gonzalez, Woods, and
  Masters]{gonzalez2009digital}
Rafael~C Gonzalez, Richard~E Woods, and Barry~R Masters.
\newblock Digital image processing, 2009.

\bibitem[Gupta(2021)]{9543483}
Bhisham~C. Gupta.
\newblock \emph{Sampling Methods}, pages 89--121.
\newblock 2021.
\newblock \doi{10.1002/9781119671718.ch4}.

\bibitem[Gustineli(2022)]{arxiv.2204.02921}
Murilo Gustineli.
\newblock A survey on recently proposed activation functions for deep learning,
  2022.
\newblock URL \url{https://arxiv.org/abs/2204.02921}.

\bibitem[Hui and Belkin(2020)]{arxiv.2006.07322}
Like Hui and Mikhail Belkin.
\newblock Evaluation of neural architectures trained with square loss vs
  cross-entropy in classification tasks, 2020.
\newblock URL \url{https://arxiv.org/abs/2006.07322}.

\bibitem[Hussain et~al.(2020)Hussain, AL-Khafaji, and Siddeq]{Hussain_2020}
Abdullah~A. Hussain, Ghadah~K. AL-Khafaji, and Mohammed~M. Siddeq.
\newblock Developed {JPEG} algorithm applied in image compression.
\newblock \emph{{IOP} Conference Series: Materials Science and Engineering},
  928\penalty0 (3):\penalty0 032006, nov 2020.
\newblock \doi{10.1088/1757-899x/928/3/032006}.
\newblock URL \url{https://doi.org/10.1088/1757-899x/928/3/032006}.

\bibitem[Kingma and Ba(2014)]{arxiv.1412.6980}
Diederik~P. Kingma and Jimmy Ba.
\newblock Adam: A method for stochastic optimization, 2014.
\newblock URL \url{https://arxiv.org/abs/1412.6980}.

\bibitem[Kundu et~al.(2020)Kundu, Mostafa, Sridhar, and
  Sundaresan]{arxiv.2012.09904}
Souvik Kundu, Hesham Mostafa, Sharath~Nittur Sridhar, and Sairam Sundaresan.
\newblock Attention-based image upsampling, 2020.
\newblock URL \url{https://arxiv.org/abs/2012.09904}.

\bibitem[Leprince et~al.(2007)Leprince, Barbot, Ayoub, and
  Avouac]{leprince2007automatic}
Sbastien Leprince, Sylvain Barbot, Franois Ayoub, and Jean-Philippe Avouac.
\newblock Automatic and precise orthorectification, coregistration, and
  subpixel correlation of satellite images, application to ground deformation
  measurements.
\newblock \emph{IEEE Transactions on Geoscience and Remote Sensing},
  45\penalty0 (6):\penalty0 1529--1558, 2007.

\bibitem[Li et~al.(2020)Li, Wang, Xie, and Ma]{li_2020}
Jianwei Li, Yongtao Wang, Haihua Xie, and Kai-Kuang Ma.
\newblock Learning a single model with a wide range of quality factors for jpeg
  image artifacts removal.
\newblock \emph{IEEE Transactions on Image Processing}, 29:\penalty0
  8842–8854, 2020.
\newblock ISSN 1941-0042.
\newblock \doi{10.1109/tip.2020.3020389}.
\newblock URL \url{http://dx.doi.org/10.1109/TIP.2020.3020389}.

\bibitem[Li and Lo(2019)]{li2019joint}
Peiya Li and Kwok-Tung Lo.
\newblock Joint image encryption and compression schemes based on 16 * 16 dct.
\newblock \emph{Journal of Visual Communication and Image Representation},
  58:\penalty0 12--24, 2019.

\bibitem[Pielawski and Wählby(2020)]{pielawski_introducing_2020}
Nicolas Pielawski and Carolina Wählby.
\newblock Introducing {Hann} windows for reducing edge-effects in patch-based
  image segmentation.
\newblock \emph{PLOS ONE}, 15\penalty0 (3):\penalty0 e0229839, March 2020.
\newblock ISSN 1932-6203.
\newblock \doi{10.1371/journal.pone.0229839}.
\newblock URL \url{https://dx.plos.org/10.1371/journal.pone.0229839}.

\bibitem[Podder et~al.(2014)Podder, Zaman~Khan, Haque~Khan, and
  Muktadir~Rahman]{podder_comparative_2014}
Prajoy Podder, Tanvir Zaman~Khan, Mamdudul Haque~Khan, and M.~Muktadir~Rahman.
\newblock Comparative performance analysis of hamming, hanning and blackman
  window.
\newblock \emph{International Journal of Computer Applications}, 96\penalty0
  (18):\penalty0 1--7, June 2014.
\newblock ISSN 09758887.
\newblock \doi{10.5120/16891-6927}.
\newblock URL
  \url{http://research.ijcaonline.org/volume96/number18/pxc3896927.pdf}.

\bibitem[Pouyanfar et~al.(2019)Pouyanfar, Sadiq, Yan, Tian, Tao, Reyes, Shyu,
  Chen, and Iyengar]{pouyanfar_survey_2019}
Samira Pouyanfar, Saad Sadiq, Yilin Yan, Haiman Tian, Yudong Tao, Maria~Presa
  Reyes, Mei-Ling Shyu, Shu-Ching Chen, and S.~S. Iyengar.
\newblock A survey on deep learning: algorithms, techniques, and applications.
\newblock \emph{ACM Computing Surveys}, 51\penalty0 (5):\penalty0 1--36,
  September 2019.
\newblock ISSN 0360-0300, 1557-7341.
\newblock \doi{10.1145/3234150}.
\newblock URL \url{https://dl.acm.org/doi/10.1145/3234150}.

\bibitem[Rahman et~al.(2018)Rahman, Islam, Shin, and
  Islam]{rahman_histogram_2018}
Md.~Atiqur Rahman, Syed Mohammed~Shamsul Islam, Jungpil Shin, and Md.~Rashedul
  Islam.
\newblock Histogram alternation based digital image compression using base-2
  coding.
\newblock In \emph{2018 {Digital} {Image} {Computing}: {Techniques} and
  {Applications}}, pages 1--8, Canberra, Australia, December 2018. IEEE.
\newblock ISBN 9781538666029.
\newblock \doi{10.1109/DICTA.2018.8615830}.
\newblock URL \url{https://ieeexplore.ieee.org/document/8615830/}.

\bibitem[Rasheed et~al.(2020)Rasheed, Salih, Siddeq, and
  Rodrigues]{rasheed_image_2020}
Mohammed~H. Rasheed, Omar~M. Salih, Mohammed~M. Siddeq, and Marcos~A.
  Rodrigues.
\newblock Image compression based on {2D} {Discrete} {Fourier} {Transform} and
  matrix minimization algorithm.
\newblock \emph{Array}, 6:\penalty0 100024, July 2020.
\newblock ISSN 25900056.
\newblock \doi{10.1016/j.array.2020.100024}.
\newblock URL
  \url{https://linkinghub.elsevier.com/retrieve/pii/S2590005620300096}.

\bibitem[Rice et~al.(2020)Rice, Wong, and Kolter]{arxiv.2002.11569}
Leslie Rice, Eric Wong, and J.~Zico Kolter.
\newblock Overfitting in adversarially robust deep learning, 2020.
\newblock URL \url{https://arxiv.org/abs/2002.11569}.

\bibitem[Santurkar et~al.(2017)Santurkar, Budden, and
  Shavit]{santurkar_generative_2017}
Shibani Santurkar, David Budden, and Nir Shavit.
\newblock Generative {Compression}.
\newblock \emph{arXiv:1703.01467 [cs]}, June 2017.
\newblock URL \url{http://arxiv.org/abs/1703.01467}.
\newblock arXiv: 1703.01467.

\bibitem[Santurkar et~al.(2018)Santurkar, Tsipras, Ilyas, and
  Madry]{arxiv.1805.11604}
Shibani Santurkar, Dimitris Tsipras, Andrew Ilyas, and Aleksander Madry.
\newblock How does batch normalization help optimization?, 2018.
\newblock URL \url{https://arxiv.org/abs/1805.11604}.

\bibitem[Shorten and Khoshgoftaar(2019)]{shorten_survey_2019}
Connor Shorten and Taghi~M. Khoshgoftaar.
\newblock A survey on {Image} {Data} {Augmentation} for {Deep} {Learning}.
\newblock \emph{Journal of Big Data}, 6\penalty0 (1):\penalty0 60, December
  2019.
\newblock ISSN 2196-1115.
\newblock \doi{10.1186/s40537-019-0197-0}.
\newblock URL
  \url{https://journalofbigdata.springeropen.com/articles/10.1186/s40537-019-0197-0}.

\bibitem[Siddeq and Rodrigues(2014{\natexlab{a}})]{siddeq2014novel}
MM~Siddeq and MA~Rodrigues.
\newblock A novel image compression algorithm for high resolution 3d
  reconstruction.
\newblock \emph{3D Research}, 5\penalty0 (2):\penalty0 7, 2014{\natexlab{a}}.

\bibitem[Siddeq and Rodrigues(2014{\natexlab{b}})]{siddeq2014new}
MM~Siddeq and Marcos Rodrigues.
\newblock A new 2d image compression technique for 3d surface reconstruction.
\newblock 2014{\natexlab{b}}.

\bibitem[Siddeq and Al-Khafaji(2013)]{siddeq2013applied}
Mohammed~Mustafa Siddeq and Ghadah Al-Khafaji.
\newblock Applied minimized matrix size algorithm on the transformed images by
  dct and dwt used for image compression.
\newblock \emph{International Journal of Computer Applications}, 70\penalty0
  (15), 2013.

\bibitem[SMPTE(2020)]{jpeg_future_2020}
SMPTE.
\newblock The future of the {JPEG} standard, 2020.
\newblock URL \url{https://www.smpte.org/blog/the-future-of-the-jpeg-standard}.

\bibitem[Steck and Garcia(2021)]{arxiv.2110.11402}
Harald Steck and Dario~Garcia Garcia.
\newblock On the regularization of autoencoders, 2021.
\newblock URL \url{https://arxiv.org/abs/2110.11402}.

\bibitem[Svoboda et~al.(2016)Svoboda, Hradis, Barina, and
  Zemcik]{svoboda_compression_2016}
Pavel Svoboda, Michal Hradis, David Barina, and Pavel Zemcik.
\newblock Compression artifacts removal using convolutional neural networks.
\newblock \emph{arXiv:1605.00366 [cs]}, May 2016.
\newblock URL \url{http://arxiv.org/abs/1605.00366}.
\newblock arXiv: 1605.00366.

\bibitem[W3Techs(2022)]{w3techs_usage_2022}
W3Techs.
\newblock Usage statistics of image file formats for websites, 2022.
\newblock URL \url{https://w3techs.com/technologies/overview/image_format}.

\bibitem[Yadav and Shukla(2016)]{7544814}
Sanjay Yadav and Sanyam Shukla.
\newblock Analysis of k-fold cross-validation over hold-out validation on
  colossal datasets for quality classification.
\newblock In \emph{2016 IEEE 6th International Conference on Advanced Computing
  (IACC)}, pages 78--83, 2016.
\newblock \doi{10.1109/IACC.2016.25}.

\bibitem[Yuan and Hu(2019)]{yuan2019research}
Shuyun Yuan and Jianbo Hu.
\newblock Research on image compression technology based on huffman coding.
\newblock \emph{Journal of Visual Communication and Image Representation},
  59:\penalty0 33--38, 2019.

\end{thebibliography}

\end{document}